\documentclass[preprint,3p]{elsarticle}

\usepackage{graphicx}
\usepackage{mathtools}
\usepackage{amssymb}
\usepackage{bm}
\usepackage{amsthm}
\usepackage[table]{xcolor}
\usepackage{amsmath}
\usepackage{mathrsfs}
\usepackage{lscape}
\usepackage{adjustbox}
\usepackage{dsfont}

\numberwithin{equation}{section}
\newtheorem{cor}{Corollary}
\newtheorem{thm}{Theorem}

\newcommand{\N}{{\mathbb N}}
\newcommand{\FFT}{{\mathcal {FFT}}}
\newcommand{\R}{{\mathbb R}}

\newcommand{\E}{{\mathbb E}}
\newcommand{\C}{{\mathbb C}}
\newcommand{\ML}{\mathcal{M}}
\DeclareMathOperator*{\corr}{corr}
\DeclareMathOperator*{\Normal}{Normal}

\newtheorem{remark}{Remark}

\journal{}

\begin{document}

\begin{frontmatter}

\title{A fast Fourier transform method for Mellin-type option pricing}

\author[label1]{D.J. Manuge\corref{cor1}}
\address[label1]{Global Risk Management, Scotiabank, 4 King Street West, Toronto, ON, Canada, M5H 1B6 \footnote{Disclaimer: The statements and conclusions of this paper do not necessarily reflect those of Scotiabank.}}
\ead{derek.manuge@scotiabank.com}


\author[label2]{P.T. Kim}
\address[label2]{Department of Mathematics and Statistics, University of Guelph, 50 Stone Road, Guelph, ON, Canada, N1G 2W1}
\ead{pkim@uoguelph.ca}

\begin{abstract}
Analytical pricing formulas and Greeks are obtained for European and American basket put options using Mellin transforms. We assume assets are driven by geometric Brownian motion which exhibit correlation and pay a continuous dividend rate. A novel approach to numerical Mellin inversion is achieved {\it {\it via}} the fast Fourier transform, enabling the computation of option values at equidistant log asset prices. Numerical accuracy is verified among existing methods for American call options. 
\end{abstract}


\end{frontmatter}


\section{Introduction}
\label{I}

An {\sl option} is a financial contract that presents its holder with the right, but not the obligation, to buy ({\sl call}) or sell ({\sl put}) a given amount of asset at some future date. In practice, the underlying asset is often the price of a stock, commodity, foreign exchange  rate, financial index or futures contract. Although many {\sl styles} of options exist, we are concerned with the valuation of {\sl European} and {\sl American} varieties. {American} options may be exercised at any time $t<T$, while {European} options can only be exercised at time $T$. In both cases, their definitions can be extended to {\sl basket} options, which differ by their dependence on $n \in \N$ underlying assets. 

Since the seminal paper of \cite{black-scholes}, much of the literature assumes assets are driven by geometric Brownian motion (GBM). Under this assumption, European option valuation relies on solving the Black-Scholes partial differential equation (PDE). With American options, the early-exercise condition gives rise to a {\sl free boundary}, in which no closed-form solution exists. The corresponding PDE is given by the inhomogeneous Black-Scholes equation as in \cite{jarrow, jacka, kimam}, where integral-based solutions are obtained. However, using the Mellin transform to solve the PDE has only recently been considered. The novelty of the Mellin transform is threefold; one, the technique requires no change of variables or reduction to a diffusion equation; two, it enables option formulas to be expressed in terms of market asset prices, rather than logarithmic asset prices; and three, there exists a numerically fast scheme to compute multi-asset option prices. For pricing financial derivatives, the Mellin technique was first introduced in \cite{zbMATH01780547}, where the authors consider the European call option. Thereafter it was implemented in \cite{PaniniS04, panini2004option}, where the authors provide solutions for European, American, and basket options on $n=2$ underlying assets. For European options, weak payoff functions \cite{CompanyJRV07}, discrete dividends \cite{comp}, transaction costs \cite{Cheng11}, and the Black-Scholes matrix equation \cite{CortesJSS05} have since been considered in detail. However, in all of these cases continuous dividends are omitted. The single-asset case with a continuous dividend is solved in \cite{frontphd} {\it via} an approach analogous to \cite{PaniniS04}, in \cite{dufresnegar} {\it via} the discounted expectation formula for options, and \cite{Cruz-Baez, rodmam} {\it via} Mellin convolution.   For American options, the single-asset case is solved in \cite{frontphd, FrontczakS10} {\it via} an approach analogous to \cite{PaniniS04}. To the authors' knowledge, the general multi-asset formula for European and American basket options is not known. 

For multi-asset options, scaling numerical procedures to higher dimensions can pose a challenge due to the curse of dimensionality. Since assets are modelled by GBM, prices are log-normally distributed, but the sum of log-normal variables is not. In fact, the sum has no closed-form distribution function, making basket option pricing a non-trivial task. By using the Mellin transform we are able to circumvent this by replacing the distribution function with the characteristic function of the log price process. The majority of numerical pricing approaches for basket options rely on estimating analytical approximations {\it via} Monte-Carlo methods \cite{basketoption}. However, this can be computationally expensive. Since basket options are $(n+1)$-dimensional in space and time, it is important to consider the complexity of the algorithm prior to computation and is our motivation for employing the fast Fourier transform (FFT).

This paper makes three non-trivial contributions to the literature. First, we extend the existing Mellin-type pricing formulas for European, American, and more generally basket options to include $n$ assets with continuous dividend rates. Second, we obtain new expressions for the {\sl{Greeks}} of multi-asset European and American options. Third, the American put option expression is discretized, yielding a new solution to the numerical pricing problem. Computation of the solution relies on numerical Mellin inversion. In this paper, two methods to treat Mellin inversion are considered: a sine-cosine series expansion and a novel approach utilizing the FFT. Our FFT solution extends the European pricing method of \cite{HurdZ10}. 

This work is organized as follows. In section 2, we derive the European put option formula on $n$ assets. In section 3, we derive the American put option formula on $n$ assets. In section 4, the American put option formula for $n$ assets is recast into a numerical procedure involving the FFT. This proposed solution also enables the explicit pricing of European put options. In section 5, we demonstrate the accuracy of the proposed method by comparing American call option prices computed using existing methods.

\section{European Options}
\label{EO}

In this section, Mellin transforms are used to derive the formula for the price of a European basket put option where assets have a continuous dividend rate and correlation. We begin by re-deriving the single-asset case in \cite{frontphd}, followed by the general multi-asset case. 

\subsection{Integral solution on one asset}
\label{EO1}

For an option issued on a single asset, the value $V=V(S,t;K;T;\sigma;r;q)$ is dependent on an underlying asset price $0 \leq S(t) < \infty$, the exercise price $K > 0$, the maturity time $0 \leq t \leq T$, the volatility (or standard deviation) $\sigma \geq 0$ of the asset, the risk-free interest rate $r \geq 0$, and continuous dividend rate $q \geq 0$. The Black-Scholes equation for the price of a European option with dividend assets driven by geometric Brownian motion is
\begin{align} \label{bse}
\frac{\partial V}{\partial t} + (r-q)S \frac{\partial V}{\partial S} + \frac{\sigma^2 S^2}{2}  \frac{\partial^2 V}{\partial S^2} -rV=0.
\end{align}
Equation \eqref{bse} must satisfy the boundary conditions
\begin{align} \label{bc1}
V(S,T)=\theta(S)=\max(K-S)=(K-S)^+ \hspace{3mm} {\text{and}} \hspace{3mm} V(S,t) \to 0 {\text{ as }} S \to \infty .
\end{align}
Let $\ML \lbrace f(x) ; w \rbrace$ denote the Mellin transform of a function $f(x) \in \R^+$ given by,
\begin{align} \label{mel}
\hat{f}(w) := \ML \lbrace f(x) ;w \rbrace = \int_0^\infty f(x) x^{w-1} dx
\end{align}
where complex variable $w$ exists on an appropriate strip of convergence in $\C$. Conversely, the inverse Mellin transform of a function $\hat{f}(x) \in \C$ is defined by
\begin{align} \label{melinv} 
f(x) =  \ML^{-1} \lbrace \hat{f}({w}) ; {x} \rbrace = \frac{1}{2 \pi i } \displaystyle\int_{a-i\infty}^{a+i\infty} \hat{f}(w) x^{-w} dw
\end{align}
where $a \in \Re(w)$, the real part of $a \in \C$. Thus, to find the Mellin transform of the Black-Scholes equation apply \eqref{mel} to equation \eqref{bse}:
\begin{align}
\frac{\partial \hat{V}(w,t)}{\partial t}+ \bigg( \frac{\sigma^2}{2} (w^2+w) -(r-q)w -r \bigg) \hat{V}(w,t)=0.
\end{align}
By the final time condition \eqref{bc1}, the general solution becomes
\begin{align} \label{epoode}
\hat{V}(w,t)= \hat{\theta}(w) \exp \bigg(- \frac{1}{2} \sigma^2 \big( w^2 +(1-k_2)w-k_1 \big) (T-t) \bigg)
\end{align}
where $k_1={2r}/{\sigma^2}$, $k_2={2(r-q)}/{\sigma^2}$, and $\hat{\theta}(w)$ is the Mellin transform of the payoff function. Hence, by Mellin inversion we obtain an expression for the price of a European put option on one asset,
\begin{align}  \label{epo}
{V}_E^P(S,t)= \frac{1}{2 \pi i} \int_{a -i \infty}^{a+ i \infty} \hat{\theta}(w) e^{\frac{1}{2} \sigma^2 \alpha(w) (T-t)} S^{-w}dw
\end{align}
where $\alpha(w)=w^2+(1-k_2)w-k_1$ and the Mellin transform of the put payoff function is
\begin{align} \label{ppf}
\hat{\theta}(w)= \frac{K^{w+1}}{w(w+1)}
\end{align}
for $\Re(w)>0$. By setting $q=0$, \eqref{epo} reduces to $(2.1.11)$ in \cite{PaniniS04}.

\subsection{Integral solution on many assets}

For an option issued on $n$ assets, let $\bm{S}=(S_1,...,S_n)'$, $\bm{\sigma}=(\sigma_1,...,\sigma_n)'$ and $\bm{q}=(q_1,...,q_n)'$. The value $V=V(\bm{S},t;K;T;\bm{\sigma};r;\bm{q})$ is dependent on the underlying asset prices  $0 \leq{S}_i(t) < \infty$, the exercise price $K > 0$, the maturity time $0 \leq t \leq T$, the asset volatilities (or standard deviations) ${\sigma}_i \geq 0 $, the risk-free interest rate $r \geq 0$, and continuous dividend rates ${q}_i \geq 0$, $\forall i$. The assets are assumed to be driven by geometric Brownian motion,
\begin{align}
dS_i=\mu_i S_i dt + \sigma_i S_i dW_i
\end{align}
where the Wiener processes satisfy $dW_i \sim \Normal (0,dt)$ and $\corr(dW_i,dW_j)=\rho_{ij}$ for $\rho_{ij} \in [-1,1]$. The risk-neutral drift
\begin{align} \label{drift}
\mu_i = r-q_i - \frac{\sigma_i^2}{2}
\end{align} 
ensures the no-arbitrage condition holds. For multivariate Brownian motion with drift, say $\bm{X}_t$, the characteristic function $\Phi(\bm{u};t):=\exp[-t\Psi(\bm{u})]=\E[\exp(i\bm{u}'\bm{X}_t)]$ is given by the exponent
\begin{align} \label{cf}
\Psi(\bm{u}) =\frac{1}{2}  \bm{u}' \Sigma \bm{u}  -i \bm{\mu}' \bm{u}.
\end{align}
It is known under these conditions that the corresponding PDE for the price of a European basket option is the generalized Black-Scholes equation:
\begin{align} \label{gbse}
 \frac{\partial V}{\partial t} +  \frac{1}{2}  \sum\limits_{i,j=1}^n \rho_{ij} \sigma_{i} \sigma_{j} S_i S_j \frac{ \partial^2 V}{\partial S_i \partial S_j} + \sum\limits_{i=1}^n (r-q_i)S_i \frac{\partial V}{\partial S_i} -rV=0.
\end{align}
We note \eqref{gbse} must satisfy the boundary conditions
\begin{align} \label{bc2}
 V(\bm{S},T)= \theta ( \bm{S})=\big(K-\sum_{i=1}^n S_i \big)^+  \hspace{3mm} {\text{and}} \hspace{3mm}  V(\bm{S},t) \to 0 {\text{  as  }} \bm{S} \to \infty  .
\end{align}
Let $\ML \lbrace f(\bm{x}) ; \bm{w} \rbrace$ denote the multidimensional Mellin transform of a function $f(\bm{x}) \in \R^{n+}$ given by,
\begin{align} \label{meln}
\hat{f}(\bm{w}) := \ML \lbrace f(\bm{x}) ; \bm{w} \rbrace = \int_{\R^{n+}} f(\bm{x}) \bm{x}^{\bm{w}-1} d\bm{x}
\end{align}
where complex variable $\bm{w}=(w_1,...,w_n)'$ exists in an appropriate domain of convergence in $\C^{n}$. Conversely, the inverse multidimensional Mellin transform of a function $\hat{f}(\bm{w}) \in \C^n$ is defined by
\begin{align} \label{melinvn}
f(\bm{x}) = \ML^{-1} \lbrace \hat{f}(\bm{w}) ; \bm{x} \rbrace = (2 \pi i)^{-n} \displaystyle\int_\gamma \hat{f}(\bm{w}) \bm{x}^{-\bm{w}} d\bm{w}
\end{align}
where $\gamma = \overset{n}{ \underset{j=1}{ \times}} \gamma_j$ are strips in $\C^n$ defined by $\gamma_j = \{ a_j+ib_j : a_j \in \R,  b_j = \pm \infty \}$ with $a_j \in \Re(w_j)$. Thus, to find the multidimensional Mellin transform of the generalized Black-Scholes equation apply \eqref{meln} to  \eqref{gbse}:
\begin{align} 
\frac{\partial \hat{V}}{\partial t} + \frac{1}{2} \displaystyle\sum_{i,j=1}^n \rho_{ij} \sigma_i \sigma_j w_i w_j \hat{V} + \frac{1}{2} \displaystyle\sum_{i=1}^n  \sigma_i^2 w_i \hat{V} +(r-q_i) \displaystyle\sum_{i=1}^n w_i \hat{V} - r\hat{V} =0.
\end{align}
By use of \eqref{drift} and \eqref{cf} we may rearrange the expression to obtain the ordinary differential equation
\begin{align} 
\frac{d\hat{V}(\bm{w},t)}{d t}=(\Psi(\bm{w}i)+ r) \hat{V}(\bm{w},t).
\end{align}
Solving {\it via} the final time condition \eqref{bc2} yields
\begin{align} 
\hat{V}(\bm{w},t)=\hat{\theta}(\bm{w})e^{-(\Psi(\bm{w}i)+ r)(T-t)}.
\end{align}
Hence, by Mellin inversion we obtain our result.
\begin{thm} The Mellin-type formula for a European basket put option on $n$ assets is given by
\begin{align} \label{euro} 
{V}_E^P(\bm{S},t) &=\ML^{-1} \big\lbrace  \hat{\theta}(\bm{w}) \Phi(\bm{w}i, T-t) \big\rbrace e^{-r (T-t)} .
\end{align}
where $\Phi(*)$ is the characteristic function of a multivariate Brownian motion with drift and the Mellin transform of the payoff function is given by
\begin{align} \label{ppfn}
\hat{\theta}(\bm{w})= \frac{ \beta_n(\bm{w}) K^{1+ \sum \bm{w}}} {(\sum \bm{w}) (\sum \bm{w}+1)}
\end{align}
for multinomial beta function $\beta_n(\bm{w})= \prod_{j=1}^{n} \Gamma(w_{j}) / \Gamma (\sum_{i=1}^{n}w_{i})$, $\bm{w} \in \C^n$, and $\Re(\bm{w})>0$
\end{thm}
\noindent The derivation of \eqref{ppfn} proceeds as follows. Consider the following expression for the $J$-dimensional Mellin transform of the put payoff function on $J$ assets:
\begin{align}
\int_{\R^{J+}} (K- \sum_{j=1}^ J S_i)^+ \prod_{j=1}^J S_j^{w_j-1} dS_j =  \frac{ \prod_{j=1}^{J} \Gamma(w_{j})}{\Gamma (2+ \sum_{j=1}^{J}w_{j})} K^{1+ \sum_{j=1}^{J} w_{j}} .
\end{align}
When $J=1$ the expression equals \eqref{ppf} and thus holds. Assume $J=n$, then for $J=n+1$
\begin{align*}
LHS &= \int_{\R^{(n+1)+}} (K- \sum_{j=1}^ {n+1}S_i)^+ \prod_{j=1}^{n+1} S_j^{w_j-1} dS_j \\
&=  \frac{ \prod_{j=1}^{n} \Gamma(w_{j})}{\Gamma (2+ \sum_{j=1}^{n}w_{j})}  \int_0^K (K-S_{n+1})^{1+\sum_{j=1}^n w_j} S_{n+1}^{w_{n+1}-1} dS_{n+1} \\
&=  \frac{ \prod_{j=1}^{n+1} \Gamma(w_{j})}{\Gamma (2+ \sum_{j=1}^{n+1}w_{j})}  K^{1+ \sum_{j=1}^{n+1}}
\end{align*}
 from Fubini's theorem and (3.191.1) in \cite{igrad80}. The result follows from the definition of the multinomial beta function and properties of gamma functions.

\begin{remark}
An application of generalized put-call parity computes the price of a European call from a put (see \cite{genpc}).
\end{remark}

\section{American Options}
\label{AO}
 
In this section, Mellin transforms are used to derive the formula for the price of an American basket put option where assets have a continuous dividend rate and correlation. We begin by re-deriving the single-asset case in \cite{frontphd}, followed by the general multi-asset case.

\subsection{Integral solution on one asset}
\label{AO1}

As mentioned, the early exercise condition of American options produces a free boundary, which we denote by the {\sl critical asset price} $S^*(t)$. For a put, when $S > S^*$ (known as the {\sl continuation region}) it is optimal to hold the option, while when $S < S^*$ (known as the {\sl exercise region}) it is optimal to exercise the option. In order for the transition at the boundary to be smooth, the option and its gradient must be continuous. The {\sl smooth pasting conditions} supply this:
\begin{align}
 \frac{\partial V(S^*, t)}{\partial S} = -1 \hspace{3mm} {\text{and}} \hspace{3mm} \theta (S)=K-S^*.
\end{align}
The value $V=V(S,t;K;T;\sigma;r;q)$ of an American option on one asset is known to satisfy the inhomogeneous Black-Scholes equation:
\begin{align} \label{ibse}
\frac{\partial V}{\partial t} + (r-q)S \frac{\partial V}{\partial S} + \frac{\sigma^2 S^2}{2}  \frac{\partial^2 V}{\partial S^2} -rV=f
\end{align}
where the {\sl early exercise function} is
\begin{align} \label{eef}
f(S,t)=
\begin{cases}
-rK+qS , &  0 \leq S \leq S^*\\
0 ,  &    S^*(t) \leq S \leq \infty 
\end{cases}
\end{align}
and the final time condition is inherited from the European case: $V(S,T)=\theta(S)=(K-S)^+$. Furthermore, the boundary conditions imposed on \eqref{ibse} are
\begin{align} \label{bc3}
V(S,T)=\theta(S)=(K-S)^+ \hspace{3mm} {\text{and}} \hspace{3mm} V(S,t) \to 0 {\text{ as }} S \to \infty .
\end{align}
Similar to the European put case, the Mellin transform of \eqref{ibse} is given by
\begin{align} \label{amode}
\frac{\partial \hat{V}(w,t)}{\partial t}+ \bigg( \frac{\sigma^2}{2} (w^2+w) -(r-q)w -r \bigg) \hat{V}(w,t)=\hat{f}(w,t).
\end{align}
The Mellin transform of the early exercise function is
\begin{align} \label{meef}
\hat{f}(w,t) = \int_0^\infty (-rK+qS)S^{w-1} dS =-rK \frac{S^*(t)^w}{w} + q \frac{S^*(t)^{w+1}}{w+1}.
\end{align}
Solving \eqref{amode} according to \eqref{bc3} and \eqref{meef} yields
\begin{align} 
\hat{V}(w,t) = \hat{\theta}(w) e^{ \frac{1}{2} \sigma^2 \alpha(w) (T-t) } &+ \int_t^T \frac{rK}{w} S^*(s)^w e^{\frac{1}{2} \sigma^2 \alpha(w)(s-t)} ds  -  \int_t^T \frac{q}{w+1} S^*(s)^{w+1} e^{\frac{1}{2} \sigma^2 \alpha(w)(s-t)} ds
\end{align}
where $\alpha(w)$ and $\hat{\theta}(w)$ are defined in section \ref{EO1}. By Mellin inversion we obtain the price of an American option on asset driven by geometric Brownian motion:
 \begin{align} \nonumber
{V}_A^P(S,t) =\frac{1}{2 \pi i} \int_{c-i \infty}^{c+ i \infty} \hat{\theta}(w) e^{ \frac{1}{2} \sigma^2 \alpha(w) (T-t) } S^{-w} dw &+  \frac{1}{2 \pi i} \int_{a-i \infty}^{a+ i \infty} \int_t^T \frac{rK}{w} \bigg( \frac{S(t)}{S^*(s)} \bigg)^{-w } e^{\frac{1}{2} \sigma^2 \alpha(w)(s-t)} dsdw \\ \label{amer1}
&-  \frac{1}{2 \pi i} \int_{a-i \infty}^{a+ i \infty} \int_t^T \frac{qS^*(s)}{w+1} \bigg( \frac{S(t)}{S^*(s)} \bigg)^{-w} e^{\frac{1}{2} \sigma^2 \alpha(w)(s-t)} dsdw.
\end{align}
The first term is the European option formula \eqref{epo}, while the second and third term represent the contribution of the early exercise premium. Note that when $q=0$, we obtain (3.1.9) in \cite{PaniniS04}.

\subsection{Integral solution on many assets}

For multiple assets, the continuation region exists for $\sum_{i=1}^n S_i >S^*$, while the exercise region exists for $\sum_{i=1}^n S_i < {S}^*$. The value $V=V(S,t;K;T;\sigma;r;q)$ of an American option on one asset is known to satisfy the inhomogeneous generalized Black-Scholes equation:
\begin{align} \label{igbsen}
 \frac{\partial V}{\partial t} +  \frac{1}{2}  \sum\limits_{i,j=1}^n \rho_{ij} \sigma_{i} \sigma_{j} S_i S_j \frac{ \partial^2 V}{\partial S_i \partial S_j} + \sum\limits_{i=1}^n (r-q_i)S_i \frac{\partial V}{\partial S_i} -rV=f
\end{align}
where the {\sl early exercise function} is
\begin{align} \label{eefn}
f(\bm{S},t)=
\begin{cases}
-rK+\sum_{i=1}^n q_iS_i , & 0 < \sum_{i=1}^n S_i \leq S^*(t) \\
0 ,  &  S^*(t) < \sum_{i=1}^n S_i < \infty.
\end{cases}
\end{align}
Similar to the European case, the boundary conditions imposed on \eqref{igbsen} are
\begin{align} \label{bc3}
 V(\bm{S},T)=\theta (\bm{S})=\big(K-\sum_{i=1}^n S_i \big)^+ \hspace{3mm} {\text{and}} \hspace{3mm} V(\bm{S},t) \to 0 {\text{ as }} \bm{S} \to \infty .
\end{align}
The smooth pasting conditions along the boundary are
\begin{align} \label{bc4}
\frac{\partial V(\bm{S}, t)}{\partial S_i} \bigg|_{\sum_{i=1}^n S_i = S^*} = -1 \hspace{3mm} {\text{and}} \hspace{3mm} \theta (\bm{S})=K-S^*.
\end{align}
The multidimensional Mellin transform of \eqref{igbsen} is given by the expression
\begin{align} \label{igbem}
\frac{\partial \hat{V}}{\partial t} + \frac{1}{2} \displaystyle\sum_{i,j=1}^n \rho_{ij} \sigma_i \sigma_j w_i w_j \hat{V} + \frac{1}{2} \displaystyle\sum_{i=1}^n  \sigma_i^2 w_i \hat{V} +(r-q_i) \displaystyle\sum_{i=1}^n w_i \hat{V} - r\hat{V} =\hat{f}.
\end{align}
By use of \eqref{drift} and \eqref{cf} we may rearrange \eqref{igbem} to obtain the ordinary differential equation
\begin{align} 
\frac{d \hat{V}(\bm{w},t)}{d t}-(\Psi(\bm{w}i)+ r) \hat{V}(\bm{w},t)=\hat{f}(\bm{w},t).
\end{align}
Solving {\it via} the final time condition \eqref{bc4} and applying Duhamel's principle yields
\begin{align}
\hat{V}(\bm{w},t) &=\hat{\theta}(\bm{w})e^{-(\Psi(\bm{w}i)+ r)(T-t)} - \int\limits_t^T \hat{f}(\bm{w},s) e^{-(\Psi(\bm{w}i)+ r)(s-t)} ds.
\end{align}
Hence, by Mellin inversion we obtain our result.

\begin{thm} The Mellin-type formula for an American basket put option on $n$ assets is given by
\begin{align} \label{amer} 
{V}_A^P(\bm{S},t) &=e^{-r (T-t)} \ML^{-1} \Big\lbrace  \hat{\theta}(\bm{w}) \Phi(\bm{w}i, T-t) \Big\rbrace -  \ML^{-1} \Big\lbrace  \int_t^T \hat{f}(\bm{w},s) \Phi(\bm{w}i, s-t) e^{-r(s-t)}  ds  \Big\rbrace
\end{align}
where $\Phi(*)$ is the characteristic function of a multivariate Brownian motion with drift,  $\hat{\theta}(*)$ is the Mellin transform of the payoff function given by \eqref{ppfn}, and the Mellin transform of the early exercise function is given by
\begin{align} \label{meef}
\hat{f}(\bm{w},t)= \frac{\beta_n(\bm{w}) ({S}^*)^{\sum \bm{w}} }{\sum \bm{w}} \bigg[ \frac{\bm{q}'\bm{w} {S}^*}{\sum \bm{w} +1}  -rK \bigg]
\end{align}
for critical asset price $S^*(t)$, multinomial beta function $\beta_n(\bm{w})= \prod_{j=1}^{n} \Gamma(w_{j}) / \Gamma (\sum_{i=1}^{n}w_{i})$, $\bm{w} \in \C^n$, and $\Re(\bm{w})>0$.
\end{thm}
\noindent The derivation for \eqref{meef} proceeds as follows. Consider the following expression for the $J$-dimensional Mellin transform of the early exercise function on $J$ assets:
\begin{align} \nonumber
\int_{\R^{J+}} \bigg(-rK+\sum_{i=1}^J q_iS_i \bigg) \prod_{j=1}^J S_j^{w_j-1} dS_j =  \frac{ \prod_{j=1}^{J} \Gamma(w_{j})  (S^*)^{\sum_{j=1}^{J} w_{j}}}{\Gamma (1+ \sum_{j=1}^{J}w_{j})} \bigg[\frac{ S^* \sum_{j=1}^J q_j w_j}{\sum_{j=1}^J w_j+ 1 }  - rK \bigg].
\end{align}
When $J=1$ the expression equals \eqref{meef} and thus holds. Assume $J=n$, then for $J=n+1$
\begin{align*}
LHS &=  \int_{\R^{(n+1)+}} \bigg(-rK+\sum_{i=1}^{n+1} q_iS_i \bigg) \prod_{j=1}^{n+1} S_j^{w_j-1} dS_j \\
&= \frac{- rK \prod_{j=1}^n \Gamma(w_j) }{ \Gamma(1+ \sum_{j=1}^n w_j)} \int_0^{S^*} (S^*-S_{n+1})^{\sum_{j=1}^n w_j } S_{n+1}^{w_{n+1}-1} dS_{n+1} \\
&+ \frac{  \sum_{j=1}^{n+1} q_j w_j \prod_{j=1}^n \Gamma(w_j) }{ \Gamma(2+ \sum_{j=1}^n w_j)} \int_0^{S^*} (S^*-S_{n+1})^{1+\sum_{j=1}^n w_j } S_{n+1}^{w_{n+1}-1} dS_{n+1} \\
&= \frac{- rK \prod_{j=1}^{n+1} \Gamma(w_j) }{ \Gamma(1+ \sum_{j=1}^{n+1} w_j)}(S^*)^{\sum_{j=1}^{n+1}}+ \frac{  \sum_{j=1}^{n+1} q_j w_j \prod_{j=1}^{n+1} \Gamma(w_j) }{ \Gamma(2+ \sum_{j=1}^{n+1} w_j)} (S^*)^{1+\sum_{j=1}^{n+1}}
\end{align*}
from Fubini's theorem and equation (3.191.1) in \cite{igrad80}. The result follows from the definition of the multinomial beta function and properties of gamma functions.

\begin{remark}
An application of generalized put-call symmetry gives the price of an American call option from a put (see \cite{MolchanovS10}).
\end{remark}
\noindent Note that the early exercise premium only contributes to the price of the option when $\sum_{i=1}^n S_i(s) \leq S^*(s)$. Otherwise the second term of \eqref{amer} is zero. By imposing the smooth pasting conditions \eqref{bc4} on \eqref{amer}, we obtain an implicit equation describing the free boundary.
\begin{cor} The critical asset price $S^*(t)$ is given by the solution of the expression
\begin{align} \label{cap} \nonumber
K-S^*(t) &=  \frac{e^{-r (T-t)}}{2\pi i} \int_\gamma  \hat{\theta}(\bm{w}) \Phi(\bm{w}i, T-t) {\bm{S}^*(t)}^{-\bm{w}} d\bm{w}  \\
 &-   \int_\gamma  \int_t^T \hat{f}(\bm{w}, s) \Phi(\bm{w}i, s-t) e^{-r(s-t)}  {\bm{S}^*(t)}^{-\bm{w}} ds  d\bm{w} .
\end{align}
\end{cor}
\noindent The critical asset price can be obtained by solving for $S^*(t)$ where $\bm{S}^*(t)=(S_1^*,...S_n^*)$ over the space of possible prices in $ \R^{n+}$ such that $S^*=\sum_{i=1}^n S_i^*$.

\section{Option Sensitivities}
\label{Greeks}
Option {\sl sensitivities} or {\sl Greeks} describe the relationship between the value of an option and changes in one of its underlying parameters. They play a vital role for risk management and portfolio optimization, since they have the ability to describe how vulnerable an option is to a particular risk factor. They are easily obtained for European and American options by passing the appropriate derivative operator under the complex integral in \eqref{amer}. For succinctness, the variable change $\tau=T-t$ is used in some of the following expressions. The first partial derivative with respect to a given asset, Delta, is given by
\begin{flalign}
&\Delta_1   := \frac{\partial V}{\partial S_i} =  - e^{-r\tau} \ML^{-1} \Big\lbrace \frac{w_i}{S_i}  \hat{\theta}(\bm{w}) \Phi(\bm{w}i, \tau) \Big\rbrace  
  +  \ML^{-1} \Big\lbrace \frac{w_i}{S_i}  \int_0^\tau \hat{f}(\bm{w}, \tau-s) \Phi(\bm{w}i, s) e^{-rs}  ds  \Big\rbrace . \\
&{\text{The cross partial derivative with respect to two independent assets is given by}} \nonumber \\
 &\Delta_2   := \frac{\partial^2 V}{\partial S_i \partial S_j} =  - e^{-r\tau} \ML^{-1} \Big\lbrace \frac{w_i}{S_i} \frac{w_j}{S_j}  \hat{\theta}(\bm{w}) \Phi(\bm{w}i, \tau) \Big\rbrace  
+  \ML^{-1} \Big\lbrace \frac{w_i}{S_i} \frac{w_j}{S_j}  \int_0^\tau \hat{f}(\bm{w}, \tau-s) \Phi(\bm{w}i, s) e^{-rs}  ds  \Big\rbrace  . 
\end{flalign}
Gamma, the second derivative with respect to the asset price is given by
\begin{flalign} \nonumber
 \Gamma := \frac{\partial^2 V}{\partial S_i^2} =  & -e^{-r \tau} \ML^{-1} \Big\lbrace w_i(1-w_i) \hat{\theta}(\bm{w}) \Phi(\bm{w}i, \tau) S_i^{-2} \Big\rbrace \\
&   -  \ML^{-1} \Big\lbrace S^{-2}  \int_t^T w_i(1-w_i)  \hat{f}(\bm{w}, s) \Phi(\bm{w}i, s-t) e^{-r(s-t)}  ds  \Big\rbrace . \\
\shortintertext{Theta, the first partial derivative with respect to time is} \nonumber \\ \nonumber
\Theta  := -\frac{\partial V}{\partial t} =& -e^{-r(T-t)} \ML^{-1} \Big\lbrace (\Psi(\bm{w}i)+r) \hat{\theta}(\bm{w}) \Phi(\bm{w}i, T-t) \Big\rbrace \\
& + \ML^{-1} \Big\lbrace  \int_t^T  (\Psi(\bm{w}i)+r-1) \hat{f}(\bm{w}, s) \Phi(\bm{w}i, s-t) e^{-r(s-t)}  ds  \Big\rbrace . \\
\shortintertext{Rho, the first partial derivative with respect to the risk-free rate of return is given by} \nonumber \\ \nonumber
\rho  := \frac{\partial V}{\partial r} = &  -\tau e^{-r \tau} \ML^{-1} \Big\lbrace (\sum_{j=1}^n w_i-1) (T-t) \hat{\theta}(\bm{w}) \Phi(\bm{w}i, \tau) \Big\rbrace \\
&- \ML^{-1} \Big\lbrace  \int_t^T (\sum_{j=1}^n w_i-1)(s-t)  \hat{f}(\bm{w}, s) \Phi(\bm{w}i, s-t) e^{-r(s-t)}  ds  \Big\rbrace .  \\
\shortintertext{Nu, the first partial derivative with respect to volatility is given by} \nonumber \\ \nonumber
\nu := \frac{\partial V}{\partial \sigma_i} = & \tau e^{-r\tau} \ML^{-1} \Big\lbrace \big[  \frac{1}{2} \displaystyle\sum_{\substack{ i,j=1 \\ i \neq j}}^n \rho_{ij} \sigma_j w_i w_j +  \displaystyle\sum_{i=1}^n  \sigma_i w_i(w_i-1) \big]  \hat{\theta}(\bm{w}) \Phi(\bm{w}i, \tau) \Big\rbrace \\
&- \ML^{-1} \Big\lbrace \big[  \frac{1}{2} \displaystyle\sum_{\substack{ i,j=1 \\ i \neq j}}^n \rho_{ij} \sigma_j w_i w_j +  \displaystyle\sum_{i=1}^n  \sigma_i w_i(w_i-1) \big] \int_0^\tau s \hat{f}(\bm{w}, \tau-s) \Phi(\bm{w}i, s) e^{-rs}  ds  \Big\rbrace .\\
\shortintertext{ Finally, the first partial derivative with respect to the dividend rate is given by} \nonumber \\ \nonumber
\Xi  :=  \frac{\partial V}{\partial q_i} =  & -\tau  e^{-r \tau} \ML^{-1} \Big\lbrace w_i \hat{\theta}(\bm{w}) \Phi(\bm{w}i, \tau) \Big\rbrace \\
&+ \ML^{-1} \Big\lbrace  \int_t^T w_i (s-t)  \hat{f}(\bm{w}, s) \Phi(\bm{w}i, s-t) e^{-r(s-t)}  ds  \Big\rbrace.
\end{flalign}
By eliminating the second term for each Greek we obtain the corresponding European option sensitivities.  Since most payoff functions are independent of the derivative operator, these expressions also hold for many path-independent multi-asset options. The American case differs because the exercise region varies with time and depends on the payoff function. Even in the simplest case of the basket option, the Mellin transform of the early exercise function is dependent on the derivative operator and must be considered to obtain expressions for other multi-asset Greeks.

\begin{remark}
 By direct substitution of the above expressions, we may prove that $(i)$ formula \eqref{euro} is a classical solution to the European pricing problem \eqref{gbse}-\eqref{bc2} and $(ii)$ formula \eqref{amer} is a classical solution to the American pricing problem \eqref{igbsen}-\eqref{bc4}.
\end{remark}

\section{Numerical Solution using the Fast Fourier Transform }
\label{FMT}
Valuing options on $n$ underlying assets is a difficult problem due to the curse of dimensionality. The issues stem from multiple integration, where the order of complexity does not scale linearly as $n$ increases. The fast Fourier transform (FFT), a numerically efficient discrete Fourier transform, is able to circumvent this problem by reducing the number of floating point operations from $O(N^{2n})$ to $O(N^n \log_2 N^n)$ (when the number of transformed points $N$ are equal across dimensions). In this section we present a new FFT-based method that enables the pricing of both European and American basket options. Recall \eqref{amer}, for strip of convergence $\gamma=\bm{a}+i\bm{b}$ where $\bm{b} \to \pm \infty$,
\begin{align} \nonumber
{V}(\bm{S},t)  &=\frac{e^{-r(T-t)}}{(2 \pi i)^{n}} \lim_{\bm{b} \to \infty} \int_{\bm{a}-i\bm{b}}^{\bm{a}+i\bm{b}} \hat{\theta}(\bm{w}) \Phi(\bm{w}i, T-t) \bm{S}^{-\bm{w}} d\bm{w} \\
&-  (2 \pi i)^{-n}   \lim_{\bm{b} \to \infty} \int_{\bm{a}-i\bm{b}}^{\bm{a}+i\bm{b}}\int_t^T \hat{f}(\bm{w},s)  \Phi(\bm{w}i, s-t) e^{-r(s-t)} \bm{S}^{-\bm{w}} ds d\bm{w} .
\end{align}
Make a change of variables by setting $\bm{w}=\bm{a}+i \bm{b}$ so that $d\bm{w}=id\bm{b}$. Then,
\begin{align} \nonumber
{V}(\bm{S},t) &=\frac{e^{-r(T-t)}}{(2 \pi)^{n}}  \lim_{\bm{b} \to \infty} \int_{-\bm{b}}^{ \bm{b}} \hat{\theta}(\bm{a}+i\bm{b}) \Phi(\bm{a}i-\bm{b}, T-t) \bm{S}^{-(\bm{a}+i\bm{b})} d\bm{b} \\
&-  (2 \pi)^{-n}  \lim_{\bm{b} \to \infty} \int_{-\bm{b}}^{\bm{b}}\int_t^T \hat{f} (\bm{a}+i\bm{b},s)  \Phi(\bm{a}i-\bm{b}, s-t) e^{-r(s-t)} \bm{S}^{-(\bm{a}+i\bm{b})} ds d\bm{b}.
\end{align}
Induce the time change $\tau=T-t$ and discretize the integrals over $\bm{b}$ and $\bm{s}$ by invoking the Trapezoid rule.
\begin{align} \nonumber
V(\bm{S},\tau) \simeq &   \frac{\Delta_b e^{-r\tau}}{  (2 \pi) ^{n}} \sum\limits_{j_1,...,j_n=0}^{N-1}  \hat{\theta}(\bm{a}+i\bm{b}_j) \Phi(\bm{a}i-\bm{b}_j, \tau)  e^{-(\bm{a}+i\bm{b}_j)'\ln(\bm{S})}  \\ \label{tr}
&  - \frac{\Delta_b \Delta_\tau}{ (2 \pi) ^{n}}  \sum\limits_{j_1,...,j_n=0}^{N-1}  \sum\limits_{l=0}^{M-1} \hat{f}(\bm{a}+i\bm{b}_j,\tau-t_l)   \Phi(\bm{a}i-\bm{b}_j, t_l)  e^{-rt_l-(\bm{a}+i\bm{b}_j)'\ln(\bm{S})}.
\end{align}
Time is parameterized  by $\bm{t}_l:=l\tau/(M-1)$ for stepsize $\Delta_\tau=\tau/M$ and vector ${l}=0,...,M-1$. Similarly, the Mellin integrals are defined by $\bm{b}_j=(b_{j_1},...,b_{j_n})$ where $b_{j_i} :=(j_i - \frac{N}{2}) \Delta_i$ for $j_i=0,...,N-1$, and $\Delta_b=\prod_{i=1}^n \Delta_i$. Hence, the multiple integral in $\bm{b}$ is approximated by a multiple sum over the lattice,
\begin{align} \nonumber
{\mathscr{B}}= \{ \bm{b}_j= (b_{j_1},...,b_{j_n}) |  \bm{j}=(j_1,...,j_n) \in \{ 0,...,N-1 \} ^n \}.
\end{align}
To evaluate the price inputs, define the initial log asset prices by the reciprocal lattice
\begin{align} \nonumber
{\mathscr{S}}=\{ \bm{s}_k= (s_{k_1},...,s_{k_n}) |  \bm{k}=(k_1,...,k_n) \in \{ 0,...,N-1 \} ^n \}
\end{align}
where $\bm{s}_k:=(k_i - \frac{N}{2}) \lambda_i$ for $k_i=0,...,N-1$. The well-known European FFT procedures of \cite{carr_madan_1999, Dempster} most noticeably differ from our approach by using log exercise prices rather than log asset prices for the FFT grid. The idea of using log-asset prices comes from \cite{HurdZ10}. Although the integral extension from European to American options is quite natural, existing FFT-based algorithms do not rely on approximating integral solutions. Rather, there are two main approaches. One approach is the FFT convolution method for Bermudan options in \cite{13284}. Bermudan options are able to provide an approximation to American options when the number of early exercise points reach infinity. The other approach is based on the linear complementarity formulation of American options. Coined the Fourier time-stepping method, the method relies on enforcing the condition $V(\bm{S},t) \geq V(\bm{S},T)$ at each timestep over the lifetime of the option \cite{Jackson}. While both of these methods sufficiently price American options, the decomposition derived in \eqref{tr} allow us to price both European and American options by considering a single formula. Further simplification can be made by recognizing that each sum in $j_i$ is truncated to $N$ evaluation points. By setting $\Delta_i \lambda_i = 2 \pi / N$  one obtains
\begin{align} \nonumber
V(\bm{S},\tau)  \simeq &  \frac{(-1)^{\sum \bm{k}}\Delta_b e^{-r\tau}}{  (2 \pi) ^{n}} \sum\limits_{j_1,...,j_n=0}^{N-1} \zeta_E  e^{-\bm{a}' \bm{s}} e^{\frac{-2 \pi i}{N} \bm{j}' \bm{k}} \\ \label{hurd} 
-& \frac{ (-1)^{\sum \bm{k}}\Delta_b \Delta_\tau}{ (2 \pi) ^{n}}  \sum\limits_{j_1,...,j_n=0}^{N-1}  \sum\limits_{l=0}^{M-1} \zeta_{EEP} e^{-rt_l-\bm{a}' \bm{s}} e^{\frac{-2 \pi i}{N} \bm{j}' \bm{k}}  
\shortintertext{where} 
&\zeta_E(\bm{b}_j)=  (-1)^{\sum \bm{j}}  \hat{\theta}(\bm{a}+i\bm{b}_j) \Phi(\bm{a}i-\bm{b}_j, \tau) e^{-r\tau}
\shortintertext{and} 
&\zeta_{EEP} (\bm{b}_j, \bm{t}_l)= (-1)^{\sum \bm{j}} \hat{f}(\bm{a}+i\bm{b}_j,\tau-t_l)   \Phi(\bm{a}i-\bm{b}_j, t_l)  e^{-rt_l}.
\end{align}
Under a change of variables $u=wi$,  \eqref{hurd} is equivalent to the method of \cite{HurdZ10} when solving for European options of unit exercise price. To obtain American options, two FFT procedures must be computed with input arrays $\zeta_E(\bm{b}_j)$ and $\zeta_{EEP}(\bm{b}_j,\bm{t}_l)$. Alternatively, by combining the integrands we need only compute one FFT, thus reducing the speed of the algorithm. An improvement in accuracy can be made by introducing the composite Simpson's rule over $\bm{k}$ and $\bm{j}$. This allows the integrand to be approximated using quadratic polynomials rather than line segments. By defining $\alpha=(3+(-1)^{1+\sum \bm{j}} - \delta_{\sum \bm{j}} )/3$, this weighted smoothing implies
\begin{align} \label{fmt}
V_A^P(\bm{S},\tau) & \simeq \frac{(-1)^{\sum \bm{k}}\Delta_b}{ (2 \pi) ^{n}} \FFT \big\lbrace \alpha \zeta_E(\bm{b}_j) \big\rbrace  e^{-\bm{a}'  \bm{s}} - \frac{ (-1)^{\sum \bm{k}}\Delta_b \Delta_\tau}{ (2 \pi) ^{n}}  \FFT \bigg\lbrace \sum\limits_{l=0}^{M-1} \alpha  \zeta_{EEP}(\bm{b}_j,\bm{t_l})  \bigg\rbrace e^{-\bm{a}' \bm{s}} \\ \label{fmt2}
& =\frac{(-1)^{\sum \bm{k}}\Delta_b}{ (2 \pi) ^{n}} \FFT  \bigg\lbrace \alpha \zeta_E(\bm{b}_j) - \alpha \Delta_\tau \sum\limits_{l=0}^{M-1}  \zeta_{EEP}(\bm{b}_j, \bm{t}_l)   \bigg\rbrace e^{-\bm{a}' \bm{s}} 
\end{align}
where the Kronecker delta function $\delta_{\sum \bm{j}}=1$ for $\sum \bm{j}=0$ and zero otherwise. The first term of \eqref{fmt} computes the price of a European put option, while the second corresponds to the early exercise premium. Evaluating both terms, or equivalently \eqref{fmt2}, retrieves the value of an American put option. The error of the numerical procedure will depend highly on the choice of $N$, $M$, $\Delta_i$ (or $\lambda_i$), and $\bm{a}$. Careful selection must be made with $\Delta_i$ (or $\lambda_i$) for the reciprical FFT grid to land on the initial log asset price specified at input.  One way this can be achieved is by solving for the root of $f(\lambda_i)=\ln(S_i)-(k_i-N/2) \lambda_i$, satisfying $0 \leq \lambda_i \leq 1$ for some $0 \leq k_i \leq N-1$. Since the size of the grid step shrinks as $k_i$ increases, $\lambda_i$ must be large enough so that the log price is contained by the range of the FFT grid, yet small enough to obtain a fine grid between log prices. A fine grid may also be achieved by increasing the number of evaluation points $N$. Further details on parameter selection and computational error with FFT-based pricing are given in \cite{HurdZ10}.

\section{Application: Pricing American Call Options}
\label{NR}
In this section we explicitly compute American call options using \eqref{fmt} and applying put-call symmetry:  $V^C_A(S,K,r,q,t) =V^P_A(K,S,q,r,t)$. Numerical methods are coded in {\tt R}. Experiments are run on a Windows 7 OS machine in {\tt R} Studio with Intel Core i3 CPU @ 2.53 GHz and 4 GB RAM. Prior to computing \eqref{fmt}, the critical asset price $S^*$  must be determined. A typical procedure is to recursively solve \eqref{cap} for ${S}^*$ at each timestep $t_l \in [0,T]$. If the parameters $\{ K,r,q,\sigma,T\}$ of an American option are known prior to pricing, the critical asset price can be calculated {\it ex-ante} and stored to reduce runtime. In practice, parameters such as volatility and time to maturity continuously change. Hence, one may wish to reduce runtime by computing an analytical approximation; often posed as an implicit function of $S^*$. We consider proposition 5.3.3. of \cite{frontphd, FrontczakS10} for a dividend-paying asset:
\begin{align} \label{CAPf}
S^*(t)=K \frac{r}{\sigma \sqrt{\delta}} \frac{2 N(\sqrt{\delta (T-t)})-1}{e^{q(T-t)} [N(\kappa)-N(\sqrt{(\delta-2q)(T-t)})]+\omega+\frac{1}{2}}
\end{align}
where
\begin{align}
\omega &= \frac{2q+\sigma \sqrt{\delta-2q}}{2\sigma \sqrt{\delta}} [2N(\sqrt{\delta(T-t)})-1] \\
\delta &= \frac{\sigma}{2} + \frac{q-r}{\sigma}+2r \\
\kappa &=\frac{\ln(S^*/K) + (r-q+\sigma^2/2)(T-t)}{\sigma \sqrt{T-t}} 
\end{align}
and $N(\cdot)$ is the cumulative Normal distribution. For our proposed experiments, prices are computed using equation \eqref{CAPf} for the critical asset price. The implicit function is numerically solved by Brent's method. 

In table \ref{amt} existing methods for computing 6-month American call option prices are compared against the benchmark binomial options formula with 10000 timesteps (True) in \cite{coxrossrub}. Including the proposed FFT method of section \ref{FMT} (FFT), the following numerical procedures are considered: the method of Barone-Adesi and Whaley (BAW) \cite{BAW}, the four-point method of Geske and Johnson (GJ4) \cite{GJ4}, the modified two-point Geske-Johnson approach of Bunch and Johnson (BJ2) \cite{BJ2}, the four-point schemes of Huang et al. (HSY4) \cite{HSY4}, the lower and upper bound approximation of Broadie and Detemple (LUBA) \cite{LUBA}, the four-point randomization method of Carr (RAN4) \cite{RAN4}, the three-point multi-piece exponential boundary approximation of Ju (EXP3) \cite{EXP3}, an approximation of Ju and Zhong (JZ) \cite{JZ}, the Gauss-Laguerre quadrature method of Frontczak and Schobel (GL) \cite{FrontczakS10}, and the Mellin-inversion scheme of Dishon and Weiss (DW) (see Appendix A). 

\begin{table}[h]
 \centering
\scalebox{0.8}{
    \begin{tabular}{cccccccccccccc} \hline
{\bf S}    &   {\bf True} &   {\bf FFT}  &  {\bf DW}  &  {\bf BAW}  &  {\bf GJ4}  &    {\bf BJ2}  &   {\bf HSY4} &  {\bf LUBA} &  {\bf RAN4} &  {\bf EXP3} &  {\bf JZ}   & {\bf GL}   \\ \hline
80&0.2194&0.2198&0.2198&0.2300&0.2191&0.2186&0.2199&0.2195&0.2188&0.2196&0.2216&0.2185 \\
90&1.3864&1.3894&1.3895&1.4050&1.3849&1.3818&1.3898&1.3862&1.3802&1.3872&1.3857&1.3851\\
100&4.7825&4.7942&4.7943&4.7821&4.7851&4.7862&4.8044&4.7821&4.7728&4.7837&4.7682&4.7835\\
110&11.0978&11.1269&11.1270&11.0409&11.0889&11.2553&11.0686&11.0976&11.0893&11.0993&11.0794&11.1120\\
120&20.0004&20.0594&20.0591&20.0000&20.0073&20.0000&20.0531&20.0000&20.0000&20.0005&20.0000&20.0000\\  \hline
80&2.6889&2.6921&2.6921&2.7108&2.6864&2.6827&2.6897&2.6893&2.6787&2.6899&2.6871&2.6788\\
90&5.7223&5.7298&5.7297&5.7416&5.7212&5.7163&5.7361&5.7231&5.7113&5.7237&5.7110&5.7195\\
100&10.2385&10.2539&10.2538&10.2417&10.2451&10.2351&10.2752&10.2402&10.2205&10.2404&10.2143&10.2265\\
110&16.1812&16.2076&16.2074&16.1520&16.1831&16.2107&16.2012&16.1817&16.1629&16.1831&16.1456&16.1756\\
120&23.3598&23.4013&23.4010&23.2883&23.3419&23.4771&23.3288&23.3574&23.3389&23.3622&23.3211&23.3828\\ \hline 
80&1.6644&1.6643&1.6644&1.6645&1.6644&1.6644&1.6644&1.6644&1.6604&1.6644&1.6644&1.6644\\
90&4.4947&4.4946&4.4947&4.4950&4.4946&4.4947&4.4947&4.4947&4.4959&4.4947&4.4947&4.4947\\
100&9.2504&9.2505&9.2506&9.2513&9.25091&9.2506&9.2506&9.2506&9.2513&9.2506&9.2507&9.2506\\
110&15.7977&15.7974&15.7975&15.7988&15.7973&15.7975&15.7975&15.7975&15.7994&15.7975&15.7977&15.7980\\ 
120&23.7061&23.706&23.7062&23.07086&23.7082&23.7062&23.7062&23.7062&23.7027&23.7062&23.7066&23.7060 \\ \hline
    \end{tabular}
}
\caption{American call option prices calculated using twelve different pricing methods at varying risk-free rate $r$, dividend rate $q$, and volatility $\sigma$. All options have a 6-month expiry and are calculated with exercise price $K=100$ for asset prices $S=\{80,90,100,110,120\}$. The first grouping is calculated with $r=0.03$, $q=0.07$, and $\sigma=0.2$. The second grouping is calculated with $r=0.03$, $q=0.07$, and $\sigma=0.4$. The third grouping is calculated with $r=0.07$, $q=0.03$, and $\sigma=0.3$.  }
\label{amt}
\end{table}

Parameter selection for each method coincides with the original references, excluding the Mellin-based FFT and DW methods which were introduced here. As previously mentioned, the spacing for the FFT grid $\Delta_b$ must be chosen {\it a priori} for the panel of option prices to land on the appropriate asset price. Since we are pricing call options by pull-call symmetry, our grid spacing will depend on $K$ instead of $S$. By fixing $N=2^{14}$ evaluation points and $K=100$ across all experiments, solving for the root of $f(\lambda)=\ln(K)-(k-N/2) \lambda$ yields a grid spacing of $\Delta_b=0.2499913$. Although the strip of convergence exists for $\Re(w)=a>0$, it must evaluated at a given point. The integrand of \eqref{amer1} tends to oscillate as $a$ approaches the endpoints on $(0,\infty)$. For this reason, the arbitrary selection of $a=1$ is made. In addition, $M=250$ timesteps is chosen to evaluate the trapezoid rule in the Mellin transform of the early exercise function. For the DW experiments we adopt $N=250$ evaluation points, $a=1$ for the strip of convergence, $M=250$ timesteps for the Mellin transform of the early exercise function, and $L=10$ for the bounds on the log-price range.

Note that we may alternatively obtain the American option price by directly computing equation \eqref{tr}. In this case, pricing error is comparable to the FFT method; the benefit of the FFT method stems from its speed, not necessarily an improvement in accuracy. Computationally, the FFT method most notably differs by generating $2^N$ option prices, while the trapezoid rule generates one. It is often the case that one wishes to determine a single option price, however computing a panel of option prices may be viable when the initial asset price is unknown. For example, suppose a stock option is issued at some future date. By forecasting an expected price range for the stock on the date of issuance, one can determine the corresponding price range for the option. This eliminates having to compute multiple valuations at different forecasted asset prices. Even if one option price is required, the FFT algorithm along with a simple index search returns the required option price in less runtime than computing the equivalent trapezoid rule. For example, using the same parameters as in table \ref{amt} the FFT method takes $\sim$5.7 seconds to run, compared with $\sim$6.5 seconds for equation \eqref{tr}. As expected, this computational efficiency is augmented as we increase dimensions.

As such, our results indicate that the proposed FFT pricing method provides accurate American call option prices. Due to the computational advantages of implementing the FFT, we consider it to be a viable alternative to existing methods. From a practical standpoint, improvements in speed can be achieved by storing the critical asset price at given parameter sets $\{ K,r,q,\sigma,\tau \}$ prior to pricing. Or as mentioned, if the tradeoff in error is warranted, one may compute an exact form for the critical asset price. Other analytical approximations may also be explored. Although we concern ourselves with American option pricing, European options are easily obtained by computing the first term in \eqref{fmt}. Since no free boundary exists in this case, there will be less error in the option price. For example, using the same parameter choices as row 1 of table \ref{amt}, the absolute pricing error is on the order of $10^{-14}$ when compared against the Black-Scholes formula. We should note that the small error may be the result of precision in {\tt R} and is well within tolerances required by practitioners. Although omitted from this manuscript, numerically pricing higher dimensional European, American, and exotic options is feasible by the proposed method.

\section{Conclusion}

In the context of Mellin transforms, we obtain analytic solutions for the fair value of basket put options and Greeks on $n$ assets with continuous dividend rates and correlation. Solutions are obtained for both European and American option styles. By expanding on the European framework of \cite{HurdZ10}, we obtain a numerical solution to the American basket put option {\it via} the fast Fourier transform. The decomposition of the solution enables the direct computation of either European or American basket option prices. By solving for the Mellin transform of alternate payoff functions, the results presented here may be used to price more complicated multi-asset options. Numerical results are compared against twelve methods for pricing American call options, including two additional approaches to treat the Mellin inversion in our main result. The results verify the efficiency and accuracy of the proposed numerical solution. 

\section*{Acknowledgments}
This research was supported in part by the Natural Sciences and Engineering Research Council of Canada, Grant DG 46204. The first author would like to thank participants of the 4th New York Conference on Applied Mathematics at Cornell University where this research was presented.



 \bibliographystyle{elsarticle-harv}

\bibliography{sample}

\appendix

\section{Dishon and Weiss Method}

The Mellin transform is equivalent to the two-sided Laplace transform under a negative logarithmic change of variables. By exploiting this relationship, numerical Mellin inversion is possible {\it via} a series expansion of sine and cosine functions as in \cite{Dishon1978129}. By letting $S=e^{-x}$ for $-L \leq x \leq L$, equation \eqref{amer1} can be adapted to use this scheme. The value of the American option is the sum of the European option and early exercise premium given by
\begin{align}
&{V}_P^E(S,t) = \frac{e^{ax}}{2L} \hat{g}(a) + \frac{e^{ax}}{L} \sum_{j=1}^N \bigg\lbrace \Re \Big[ \hat{g}\Big(a+ \frac{\pi i j}{L}\Big) \Big] \cos \Big(\frac{\pi j x}{L} \Big) - \Im \Big[ \hat{g} \Big(a+\frac{\pi i j}{L}\Big) \Big] \sin \Big( \frac{\pi jx}{L} \Big) \bigg\rbrace
\end{align}
and
\begin{align}
&{V}_P^{EEP}(S,t) = \frac{e^{ax}}{2L} \hat{h}(a) + \frac{e^{ax}}{L} \sum_{j=1}^N \bigg\lbrace \Re \Big[ \hat{h}\Big(a+ \frac{\pi i j}{L}\Big) \Big] \cos \Big(\frac{\pi j x}{L} \Big) - \Im \Big[ \hat{h} \Big(a+\frac{\pi i j}{L}\Big) \Big] \sin \Big( \frac{\pi jx}{L} \Big) \bigg\rbrace
\end{align}
respectively. By defining the time change $\tau=T-t$ and imposing the Trapezoid rule we obtain $\hat{g}(w;\tau)=\exp(-r \tau)\hat{\theta}(w) \Phi(w i;\tau)$ and $\hat{h}(w;\tau)=\Delta_t \sum_{l=0}^{M-1} \exp(-rt_l) \hat{f}(w,\tau-t_l) \Phi(w i; t_l) $. The time integral in \eqref{amer} is approximated with truncation $M$ and stepsize $\Delta_t=\tau/M$. To achieve a faster rate of convergence, $L$ should be chosen so that $| x/L| \leq 1/2$ and when the strip of converge is finite, $a$ should be the midpoint. As before, an application of put-call symmetry yields the corresponding call price.

\end{document}